\documentclass{webofc}

\usepackage[varg]{txfonts}   
\usepackage{hyperref}
\usepackage{url}
\usepackage{float}
\hypersetup{colorlinks=true,citecolor=blue,urlcolor=blue,linkcolor=blue}
\usepackage{lineno, blindtext, indentfirst}
\begin{document}
\title{Production of Light Nuclei in Au+Au Collisions with the STAR BES-II Program}
%
%

\author{\firstname{Yixuan} \lastname{Jin}\inst{1}\fnsep\thanks{\email{jyx@mails.ccnu.edu.cn}} (For the STAR Collaboration) 
}

\institute{Institute of Particle Physics and Key Laboratory of Quark \& Lepton Physics (MOE), \\Central China Normal University, Wuhan, 430079, China.}

\abstract{
    The yields and ratios of light nuclei in heavy-ion collisions offer a method to distinguish between the thermal and coalescence models. Ratios such as $\rm N_t \times N_p/N_d^2$ and $\rm N_{^3He} \times N_p/N_d^2$ are suggested as potential probes to investigate critical phenomena within the QCD phase diagram. The significantly larger datasets from STAR BES-II compared to BES-I, combined with enhanced detector capabilities, allow for more precise measurements. In this proceeding, we present the centrality and energy dependence of transverse momentum spectra and particle yields of (anti-)proton, (anti-)deuteron, and $\rm ^3He$ at BES-II energies ($\sqrt{s_{\rm NN}}$ = $7.7 - 27$ GeV), as well as the light nuclei to proton yield ratios and coalescence parameters $(B_2(\rm d)$ and $B_3(\rm ^3He))$.
}
\maketitle
\section{Introduction}
\label{intro}
    In Quantum Chromodynamics (QCD) framework, quarks and gluons are confined within hadrons at low temperatures and baryon densities. Conversely, at higher temperatures and/or high baryon densities, hadronic matter transitions into quark-gluon plasma (QGP), where the dominant degrees of freedom are quarks and gluons. Experimentally, as the initial temperature $(T)$ and the baryon chemical potential $(\mu_B)$ vary with the centre-of-mass collision energy, the phase diagram can be explored by adjusting the beam energy \cite{PhaseScan, QCDbook2022, JHChen}.

    The Beam Energy Scan (BES) program at the Relativistic Heavy-Ion Collider (RHIC) aims to investigate QGP signatures, understand the nature of the phase transition, and identify the conjectured critical point. The second phase of the Beam Energy Scan (BES-II), with about ten times more data than the BES-I phase and enhanced detector capabilities, is driven by the need for more precise experimental results, allowing more firm physics conclusions.

    Based on the coalescence model, the light nuclei compound ratios such as $\rm N_t \times N_p/N_d^2$ and $\rm N_{^3He} \times N_p/N_d^2$ are sensitive to the neutron density fluctuations. Therefore, the ratios have been suggested as a potential probe to search for the critical phenomena in the QCD phase diagram \cite{kSun1}. The STAR BES-I and fixed-target program have measured the light nuclei production in Au+Au collisions \cite{BESI_d, BESI_t, 3GeV}.

    This proceeding presents the transverse momentum spectra, particle yields, and yield ratios for light nuclei in Au+Au collisions at $\sqrt{s_{\rm NN}}$ = $7.7 - 27$ GeV from the STAR BES-II program. Light nuclei to proton yield ratios decrease with increasing collision energy. Furthermore, the coalescence parameters $(B_2(\rm d)$ and $B_3(\rm ^3He))$ demonstrate centrality and $p_{\rm T}$ dependence, indicating the influence of collective expansion.

\section{Analysis details}
    Particles are identified using the Time Projection Chamber (TPC) at low $p_{\rm T}$, with Time of Flight (TOF) added at higher $p_{\rm T}$. After signal extraction, the spectra are corrected for energy loss, TPC tracking efficiency, TOF matching efficiency, and knock-out effects specifically for protons and deuterons. The feed-down contributions of proton and anti-proton from the weak decay of strange baryons are also applied.

    Point-to-point systematic uncertainties on the spectra are estimated by varying track selection and analysis cuts. The systematic uncertainties of $\mathrm{d}N/\mathrm{d}y$ are assessed by considering the point-to-point uncertainties in spectra, and the extrapolated range from the spectra fitting by examining the deviation between the Blast-Wave model fit and the double-$p_{\rm T}$ $(p_0 e^{-p_{\rm T}^2 / p_1^2} + p_2 e^{-p_{\rm T}^2 / p_3^2})$ function.

\section{Results}
\subsection{Corrected transverse momentum spectra}
\begin{figure}[H]
     \centering
     \includegraphics[width=6.3cm,clip]{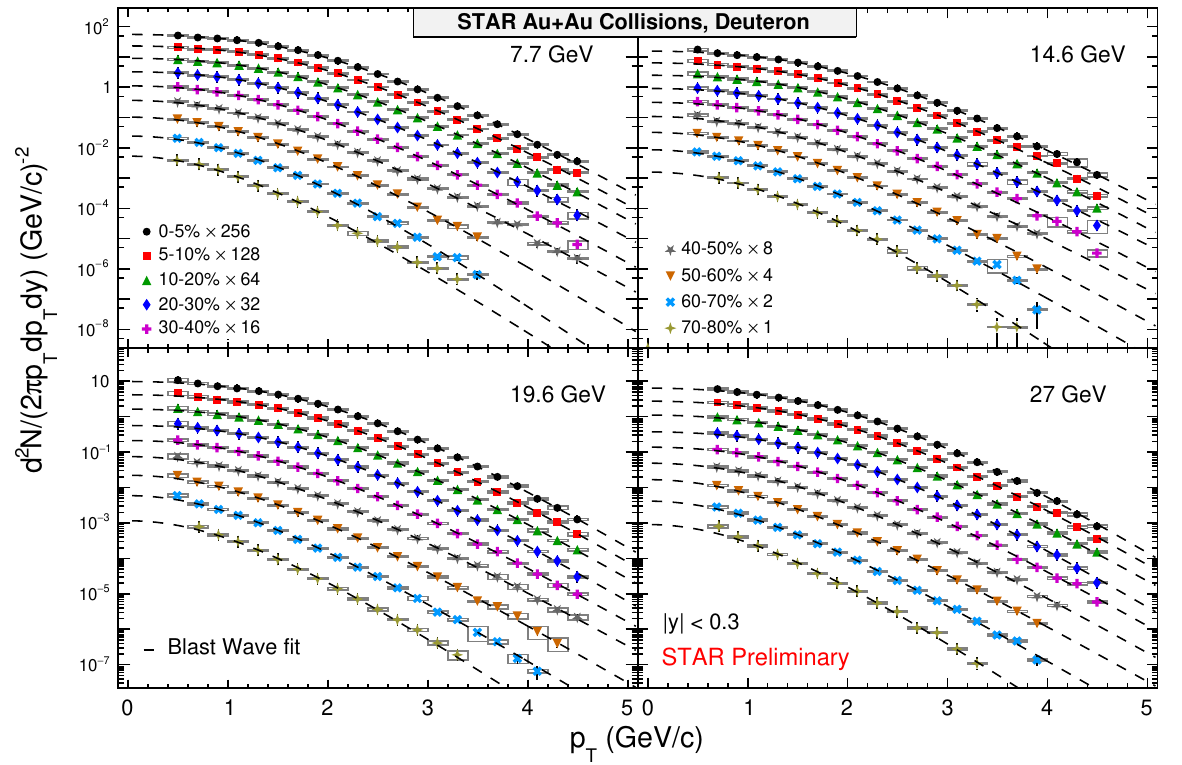}
     \includegraphics[width=6.3cm,clip]{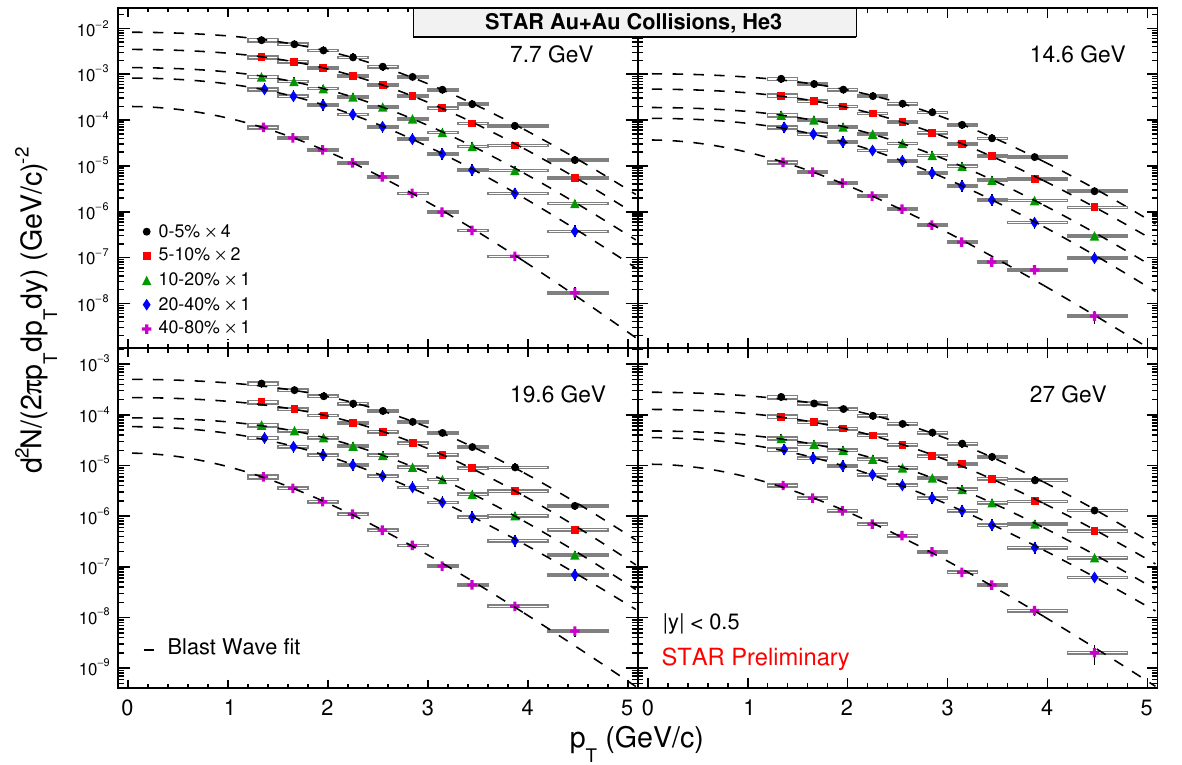}
     \caption{The transverse momentum spectra of deuteron (left) and $\rm ^3He$ (right) at mid-rapidity (with $|y| < 0.3$ for deuteron and $|y| < 0.5$ for $\rm ^3He$) for different centrality bins in Au+Au collisions from the STAR BES-II program. The dashed lines represent the corresponding Blast-Wave fits. Statistical and systematic uncertainties are indicated by vertical lines and boxes, respectively.}
     \label{fig-1}
\end{figure}

    The transverse momentum ($p_{\rm T}$) spectra of deuterons and $\rm ^3He$ in Au+Au collisions from the STAR BES-II program are shown in Fig. \ref{fig-1}. Due to higher statistics, the BES-II spectra cover more centrality bins, and remain consistent with those from BES-I within the uncertainties. Additionally, thanks to the inner TPC (iTPC), the $p_{\rm T}$ acceptance is significantly enhanced, and the extended $p_{\rm T}$ ranges in BES-II contribute to a reduction in the systematic uncertainties of the particle yields.
    
\subsection{Particle Yields}
\begin{figure}[H]
     \centering
     \includegraphics[width=6cm,clip]{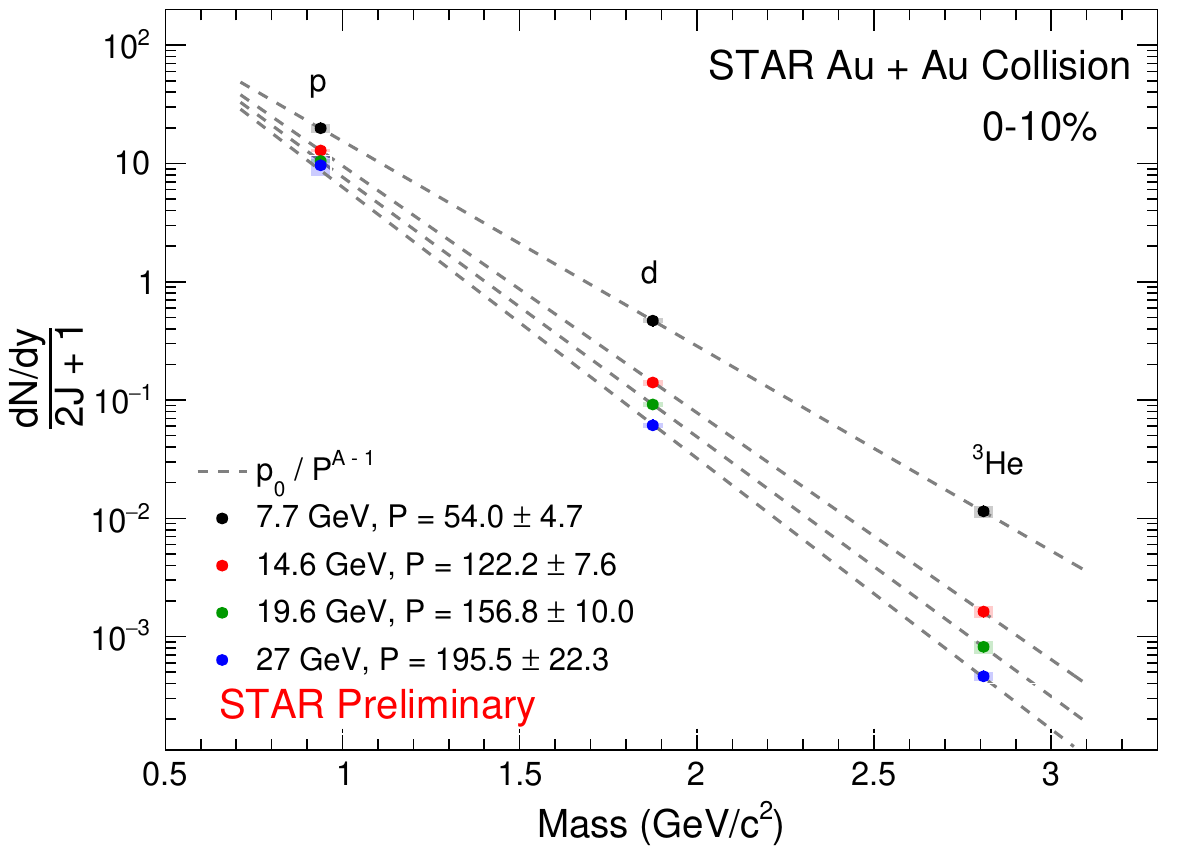}
     \caption{The mass $m_A$ (where $A$ is the mass number) dependence of $\mathrm{d}N/\mathrm{d}y$ for protons, deuterons, and $\rm ^3He$, scaled by the spin degeneracy factor $2J + 1$, is shown for Au+Au collisions at BES-II energies of $\sqrt{s_{\rm NN}} = 7.7 - 27$ GeV. Statistical uncertainties are indicated by vertical lines, while systematic uncertainties are shown by boxes. The yields are fitted using dashed lines representing exponential functions.}
     \label{fig-2}
\end{figure}

    The particle yields $(\mathrm{d}N/\mathrm{d}y)$ are extracted from $p_{\rm T}$ spectra, with measurement ranges integrated over $p_{\rm T}$ and extrapolations carried out using Blast-Wave model fits. Figure \ref{fig-2} shows $\mathrm{d}N/\mathrm{d}y$ for different particles with the scaling of their corresponding spin degeneracy factor $2J + 1$. These scaled yields of light nuclei follow an exponential distribution of the form $p_0 / P^{A - 1}$, where $P$ is the penalty factor determined by the Boltzmann factor $e^{(m_N - \mu_B) / T}$ in the thermal model. The penalty factor increases with higher beam energy, indicating that it becomes more challenging to form high-mass objects.

\subsection{Particle Yield Ratios}
\begin{figure}[H]
     \centering
     \includegraphics[width=6.4cm,clip]{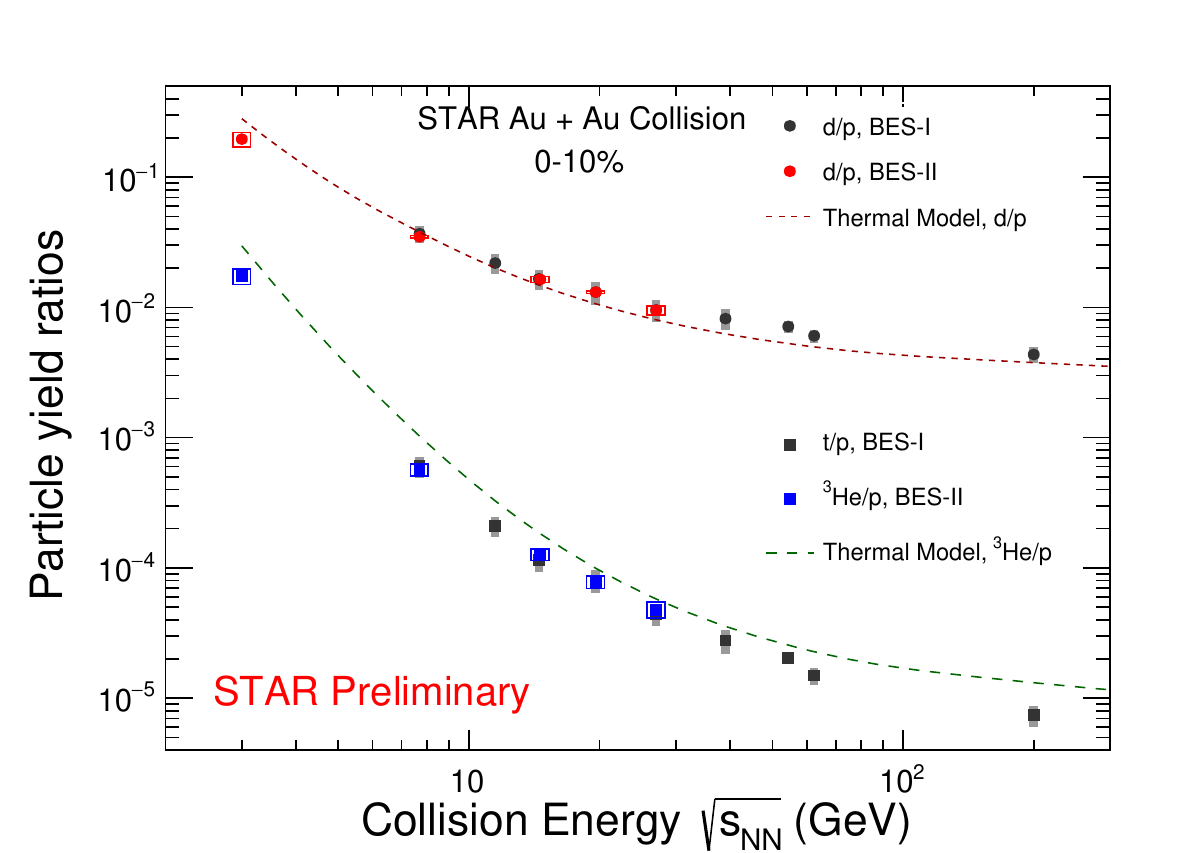}
     \caption{The energy dependence of $\rm N_d/N_p$ (the red circles) and $\rm N_{^3He}/N_p$ (the blue squares) in the most central 0-10\% collisions from BES-II. The gray points represent $\rm N_d/N_p$ and $\rm N_t/N_p$ from BES-I. Vertical lines and boxes represent statistical and systematic uncertainties, respectively. The lines show the light nuclei to proton yield ratios from the thermal model.}
     \label{fig-3}
\end{figure}

    Figure \ref{fig-3} presents the light nuclei to proton yield ratios consistent with BES-I results within uncertainties. The $\rm N_d/N_p$ and $\rm N_{^3He}/N_p$ ratios decrease monotonically with increasing collision energy, and the difference between these ratios becomes smaller at lower energies. The thermal model, shown by dashed lines, describes the $\rm N_d/N_p$ ratios well. However, it overestimates the $\rm N_t/N_p$ and $\rm N_{^3He}/N_p$ ratios by about factor 2, likely due to the effects of hadronic re-scattering during the hadronic expansion stage.

\subsection{Coalescence Parameters}
In the coalescence picture:
\begin{equation}
     E_A \frac{\mathrm{d}^3 N_A}{\mathrm{d} p_A^3} = B_A \left( E_p \frac{\mathrm{d}^3 N_p}{\mathrm{d}p_p^3} \right)^Z \left( E_n \frac{\mathrm{d}^3 N_n}{\mathrm{d}p_n^3} \right)^{A-Z} \approx B_A \left( E_p \frac{\mathrm{d}^3 N_p}{\mathrm{d}p_p^3} \right)^A
\end{equation}
    where $A$ is the mass number and $Z$ is the charge number of the light nucleus. $p_n, p_p, p_A$ are the momenta of the neutron, proton, and nucleus.

\begin{figure}[H]
     \centering
     \includegraphics[width=6.4cm,clip]{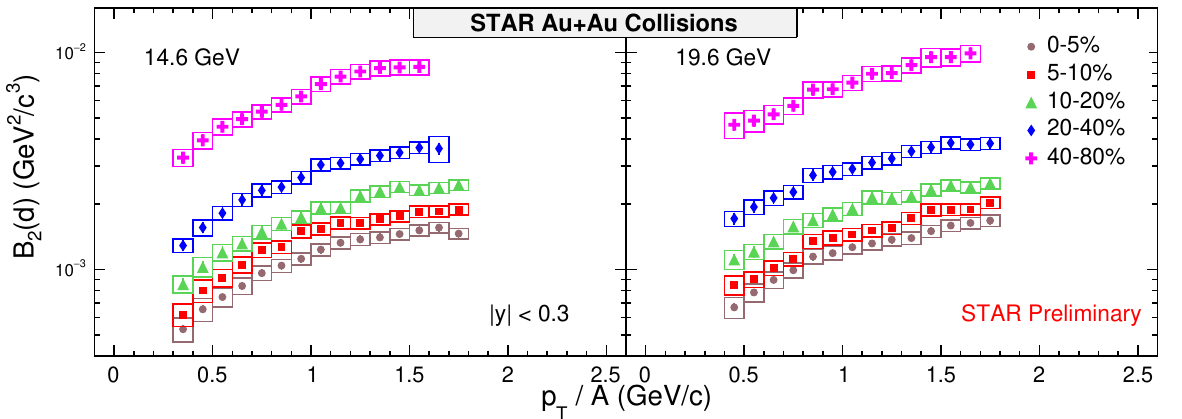}
     \includegraphics[width=6.4cm,clip]{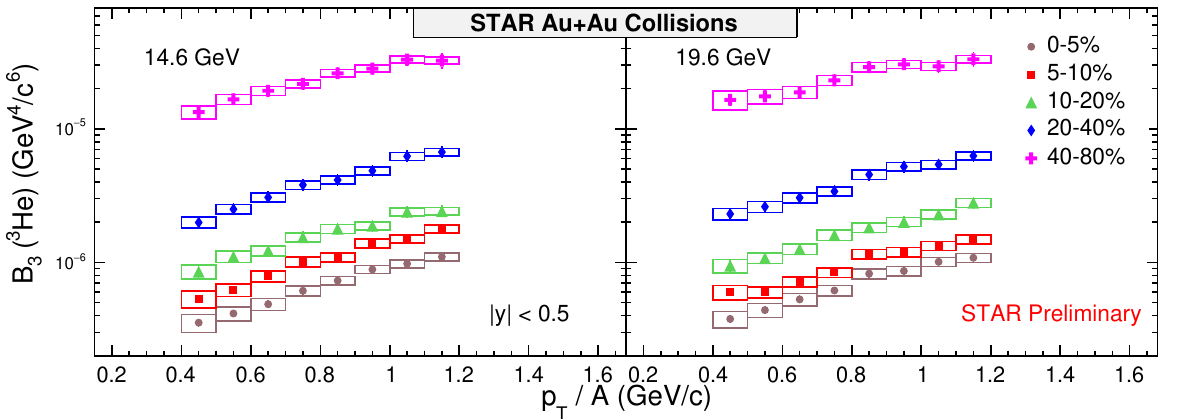}
     \caption{Number of constituent nucleon (NCN) scaling for $p_{\rm T}$ dependence of $B_2$ for deuteron (left) and $B_3$ for $\rm ^3He$ (right) in centrality 0-5\%, 5-10\%, 10-20\%, 20-40\% and 40-80\% in Au+Au collisions at $\sqrt{s_{\rm NN}}$ = 14.6 and 19.6 GeV. The boxes represent the systematic uncertainties.}
     \label{fig-4}
\end{figure}

    The invariant yield of light nuclei is proportional to the invariant yield of nucleons, with the coalescence parameter $B_A$ reflecting the probability of nucleon coalescence and relating to the local nucleon density. Figure \ref{fig-4} shows coalescence parameters as a function of $p_{\rm T} / A$ across different centralities. $B_A$ increases with transverse momentum, suggesting an expanding collision system, and also increases from central to peripheral collisions, likely due to a decreasing source volume.

\section{Conclusions}
    The production of primordial proton, deuteron, and $\rm ^3He$ in Au+Au collisions at $\sqrt{s_{\rm NN}} = 7.7 - 27$ GeV from RHIC STAR BES-II is reported in these proceedings. The particle ratios $\rm N_d / N_p$ and $\rm N_{^3He} / N_p$ show a monotonic decrease with collision energy. The thermal model over-predicts $\rm N_t / N_p$ and $\rm N_{^3He} / N_p$ by a factor of about 2. The coalescence parameter $B_A$ increase from low to high $p_{\mathrm{T}}$ due to collective expansion, and decreasing source volume results in a rise of $B_A$ from central to peripheral collisions. The analysis of BES-II data at 9.2, 11.5, and 17.3 GeV is currently underway, promising further insights into the energy-dependent behavior of the system.

\section{Acknowledgement}
    This work was supported by National Key Research and Development Program of China (No.2022YFA1604900, No.2020YFE0202002), National Natural Science Foundation of China (No.12122505) and the Fundamental Research Funds for the Central Universities (CCNU220N003).

\end{document}